\documentclass[12pt]{article}
\usepackage{epsfig,cite}
\usepackage {amsmath}
\usepackage{amssymb}
\include{epsf}
\newcount\timecount
\newcount\hours \newcount\minutes  \newcount\temp \newcount\pmhours
\hours = \time
\divide\hours by 60
\temp = \hours
\multiply\temp by 60
\minutes = \time
\advance\minutes by -\temp
\def\hour{\the\hours}
\def\minute{\ifnum\minutes<10 0\the\minutes
            \else\the\minutes\fi}
\def\clock{
\ifnum\hours=0 12:\minute\ AM
\else\ifnum\hours<12 \hour:\minute\ AM
      \else\ifnum\hours=12 12:\minute\ PM
            \else\ifnum\hours>12
                 \pmhours=\hours
                 \advance\pmhours by -12
                 \the\pmhours:\minute\ PM
                 \fi
            \fi
      \fi
\fi
}

\def\monthname{\relax\ifcase\month 0/\or January\or February\or
   March\or April\or May\or June\or July\or August\or September\or
   October\or November\or December\else\number\month/\fi}

\def\bold#1{\setbox0=\hbox{$#1$}%
     \kern-.025em\copy0\kern-\wd0
     \kern.05em\copy0\kern-\wd0
     \kern-.025em\raise.0433em\box0 }

\def\beq{\begin{equation}}
\def\eeq{\end{equation}}

\def\ga{\mathrel{\raise.3ex\hbox{$>$\kern-.75em\lower1ex\hbox{$\sim$}}}}
\def\la{\mathrel{\raise.3ex\hbox{$<$\kern-.75em\lower1ex\hbox{$\sim$}}}}
\def\gev{{\rm \, Ge\kern-0.125em V}}
\def\tev{{\rm \, Te\kern-0.125em V}}
\def\gyr{{\rm \, G\kern-0.125em yr}}

\def\gappeq{\mathrel{\rlap {\raise.5ex\hbox{$>$}}
{\lower.5ex\hbox{$\sim$}}}}
\def\lappeq{\mathrel{\rlap{\raise.5ex\hbox{$<$}}
{\lower.5ex\hbox{$\sim$}}}}
\def\Toprel#1\over#2{\mathrel{\mathop{#2}\limits^{#1}}}

\def\m12{m_{1\!/2}}

\def\bea{\begin{eqnarray}}
\def\eea{\end{eqnarray}}

\def\beq{\begin{equation}}
\def\eeq{\end{equation}}

\begin{document}

\begin{titlepage}
\pagestyle{empty}
\baselineskip=21pt
\rightline{UMN--TH--3136/13, FTPI--MINN--13/04, IPMU13-0044}
\vspace{0.2cm}
\begin{center}
{\large {\bf Universality in Pure Gravity Mediation }}
\end{center}
\vspace{0.5cm}
\begin{center}
{\bf Jason L. Evans}$^{1}$,
{\bf Masahiro Ibe}$^{2,3}$ {\bf Keith~A.~Olive}$^{1}$
and {\bf Tsutomu T. Yanagida}$^{3}$\\
\vskip 0.2in
{\small {\it
$^1${William I. Fine Theoretical Physics Institute, School of Physics and Astronomy},\\
{University of Minnesota, Minneapolis, MN 55455,\,USA}\\
$^2${ ICRR, University of Tokyo, Kashiwa 277-8582, Japan}\\
$^3${Kavli IPMU, TODIAS, University of Tokyo, Kashiwa 277-8583, Japan}\\
}}
\vspace{1cm}
{\bf Abstract}
\end{center}
\baselineskip=18pt \noindent
{\small
If low energy supersymmetry is realized in nature, the apparent discovery of a Higgs boson with mass
around $125$\,GeV suggests a supersymmetric mass spectrum in the TeV or multi-TeV range.
Multi-TeV scalar masses are a necessary component of supersymmetric models with pure
gravity mediation or in any model with strong moduli stabilization.
Here, we show that full scalar mass universality remains viable as long as the ratio of Higgs vevs,
$\tan \beta$ is relatively small ($\lesssim 2.5$).
We discuss in detail the low energy (observable) consequences of these models.
}


\vfill
\leftline{February 2013}
\end{titlepage}

\section{Introduction}

In the years leading to the operation of the LHC, one of the most studied low energy
models of supersymmetry is the so-called constrained minimal supersymmetric standard
model (CMSSM) \cite{cmssm}. In the CMSSM, soft supersymmetry-breaking scalar masses, gaugino
masses and tri-linear $A$-terms are assumed to originate at some high mass scale
and be universal at the supersymmetric grand unified theory (GUT) scale.
In addition to the sign of the Higgs
mixing mass, $\mu$, the theory has four free parameters: the universal scalar mass, $m_0$,
the universal, gaugino mass, $m_{1/2}$, the universal $A$-term, $A_0$, and the
ratio of the two Higgs vacuum expectation values, $\tan \beta$. While the CMSSM takes its
motivation from supergravity, the minimal supergravity model
\cite{Fetal,acn,bfs} (that is, with a flat K\"ahler potential),
or mSUGRA, has one fewer parameters with the relation between the bilinear supersymmetry
breaking term, $B$ and $A$ imposed at the universality scale \cite{vcmssm}. In this case, $\tan
\beta$ is no longer a free parameter and thus the theory has only three parameters and a sign.

An even simpler version of mSUGRA is known as pure gravity mediation (PGM) \cite{pgm,pgm2,ArkaniHamed:2012gw}.
In the minimal model of PGM, scalars again obtain universal scalar masses from
supersymmetry breaking. However, at the tree level, there is no source for
either gaugino masses or $A$-terms. Hence the theory is reduced to a single
parameter which may be taken as the gravitino mass, $m_{3/2}$, and $m_0 = m_{3/2}$.
At one-loop, gaugino masses and $A$-terms are generated through anomalies \cite{anom}.
Thus one expects $m_{1/2}, A_0 \ll m_0$ in these models, reminiscent of split supersymmetry
\cite{split}.

It is well known, however, that supergravity models contain one or several moduli
which can lead to a disastrous cosmology \cite{pp}. Moduli, or the Polonyi field in
particular \cite{Polonyi}, may come to dominate the energy density of the early
universe and release an excess of entropy after the period of big bang nucleosynthesis.
While many potential solutions to the Polonyi problem have been discussed
\cite{sol}, it was shown that in general models with strong moduli stabilization
\cite{klor,lmo,dlmmo,Dudas:2006gr},  which also resolve the cosmological moduli problems,
lead directly to the type of sparticle spectrum found in PGM models.

There is, of course, as yet no direct evidence for supersymmetry. The continuing absence
of supersymmetric particles at the LHC~\cite{lhc} is putting increasing
pressure on supersymmetric models such as the CMSSM \cite{postlhc}.
Furthermore, the apparent discovery by
the ATLAS and CMS experiments of a new boson consistent with
the Standard Model Higgs boson \cite{lhch} with a mass of  around $125$--$126$\,GeV
puts severe pressure on models with $\cal{O}$(100) GeV scalars\footnote{Some possible exceptions are the NMSSM and models with non-decoupling D-terms.} \cite{125-other,fp1}.
Indeed, when it was recognized that
the top squarks may increase $m_h$ to about $130$\,GeV within the
MSSM \cite{mh,oyy}, it was noted that a discovery of a Higgs boson
with a mass of about $125$\,GeV would imply a supersymmetry breaking scale of order\footnote{Scalar masses this large necessitate fine tuning in the electroweak sector.}
$100$\,TeV \cite{oyy} and limit the high scale in split supersymmetry models \cite{splitlimit}.
While possibilities for a successful phenomenology remain for the CMSSM \cite{eo6},
the data seems to point beyond the CMSSM \cite{elos}.

Lacking a rich spectrum at low energy, PGM models are fully consistent with
the current non-detection of supersymmetry at the LHC and a relatively
heavy Higgs of around $125$--$126$\,GeV\,\cite{pgm,pgm2,dlmmo}, and as noted above,
the simplest version of PGM models is based on scalar mass universality.
However, if one wants to maintain the possibility of radiative electroweak symmetry breaking (EWSB)
in PGM, at least one additional parameter is needed. As we will show, universal PGM models are quite constrained and yet the possibility of experimental detection at the LHC remains viable.

In models with scalar mass universality such as the CMSSM, mSUGRA or PGM,
if the input supersymmetry breaking scale is chosen to be the GUT scale (i.e.
the scale at which gauge coupling unification occurs), one can not choose arbitrarily large
universal scalar masses and insist on a well defined electroweak symmetry breaking vacuum
(i.e., $\mu^2 > 0$) \cite{fp1,fp2}.  Furthermore, in models based on supergravity,
a solution for $\tan \beta$ (insisting on the supergravity relation between $A_0$ and $B_0$)
becomes remote.  The latter problem can be easily circumvented by adding a single non-minimal
term to the K\"ahler potential \cite{gm,ikyy,dmmo}. The problem of obtaining
consistent solutions for the low energy electroweak vacuum however is more difficult.
One can consider 1) that the input universality scale, $M_{in}$ lies above the GUT scale
\cite{superGUT,dmmo,dlmmo}. In this case, scalar masses and couplings run between
$M_{in}$ and $M_{GUT}$ may be sufficient for producing solutions to the
Higgs minimization equations.
This produces an effective non-universality of
scalar masses at the GUT scale.
Thus, one can assume 2) that there are in effect
non-universal terms in the K\"ahler potential.
So long as non-universality is restricted
to the Higgs sector, contributions to flavor changing neutral currents will remain suppressed.

In this paper, we will show that PGM models with GUT scale universality are phenomenologically
viable for a restricted range in the two parameters which specify the model:
$\tan \beta$ and $m_{3/2}$. In section 2, we compare PGM to mSUGRA and discuss the origin of our boundary masses.  In section 3.1, we discuss the the one-loop approximation for the parameter scans.  Section 3.2 contains our parameters scans.
We show that $\tan \beta$ is restricted to a narrow range from about $1.7$--$2.5$.
The Higgs mass is found to lie in the range $m_h = 126 \pm 2$\,GeV for gravitino masses in the range
about $300$--$1500$\,TeV.
The lightest supersymmetric particle (LSP) is a wino which is nearly degenerate
with a wino-like chargino.
Due to sizable radiative corrections to the neutralino and chargino masses at low
$\tan \beta$ \cite{piercepapa}, the LSP  mass is very sensitive to the sign of $\mu$.
For $\mu > 0$, the LSP tends to be very light, and the LEP
constraint on the chargino mass \cite{LEPsusy} becomes relevant.
For $\mu < 0$, there is the distinct possibility that the wino is sufficiently
heavy that the relic density falls within the WMAP range\,\cite{wmap}.
This occurs when $m_{3/2} \simeq 460-500$ TeV, and corresponds to
a Higgs mass in the range 122 -- 126 \,GeV.
Thus, we would conclude that for $\mu > 0$,
either the dark matter comes from a source other than supersymmetry,
or winos are produced non-thermally through moduli or gravitino
decay \cite{ggw,hep-ph/9906527,Ibe:2004tg,Acharya:2008bk} or for $\mu < 0$,
a viable model with thermal wino dark matter is possible for a narrow range in $m_{3/2}$.
In section 4, we present our conclusions.

\section{PGM and Universality}
To put PGM models in perspective, we begin with a brief review of minimal supergravity (mSUGRA).
Because of its flat K\"ahler potential,  mSUGRA has a low-energy potential of the form~\cite{Fetal,acn,bfs}
 \begin{eqnarray}
V  & =  &  \left|{\partial W \over \partial \phi^i}\right|^2 +
\left( A_0 W^{(3)} + B_0 W^{(2)} + h.c.\right)  + m_{3/2}^2 \phi^i \phi_i^*  \, ,
\label{pot}
\end{eqnarray}
where the $\phi_i$'s are the low energy fields, $W$ is the low-scale superpotential,
\beq
W =  \bigl( y_e H_1 L e^c + y_d H_1 Q d^c + y_u H_2
Q u^c \bigr) +  \mu H_1 H_2  \, ,
\label{WMSSM}
\eeq
with the SU(2) indices being suppressed. Here,  $H_{1,2}$ are the Higgs doublets, and $\mu$ is their mass mixing term.  $W^{(2)}$ and $W^{(3)}$ are the supersymmetry breaking bilinear and trilinear terms, and $m_{3/2}$ is the gravitino mass. Having adopted a flat K\"ahler potential in mSUGRA, the scalar masses are universal and are proportional to the gravitino mass, $m_{3/2}$.  The universal boundary condition $m_0=m_{3/2}$ is defined at some input scale $M_{in}$, usually chosen to be the GUT scale. In contrast  to the CMSSM, the bilinear term of mSUGRA is not a free parameters but is related to the trilinear term through, $B_0=A_0-m_0$. If the gauge kinetic function, $h_{\alpha \beta} \propto \delta_{\alpha \beta}$, the gaugino masses are also universal at the input scale.

The remaining parameters, $\mu$ and $\tan\beta$, are determined by minimizing the Higgs potential which gives
\beq
\mu^2=\frac{m_1^2-m_2^2\tan^2\beta+\frac{1}{2}m_Z^2(1-\tan^2\beta)+\Delta_{\mu}^{(1)}}{\tan^2\beta-1+\Delta_{\mu}^{(2)}} \, ,
\label{eq:mu}
\eeq
and
\beq
B \mu = - \frac{1}{2}(m_1^2+m_2^2+2\mu^2)\sin 2\beta +\Delta_B \, ,
\label{eq:muB}
\eeq
where $m_{1,2}$ are the weak-scale soft masses of the of the Higgs fields.
The $\Delta_B$ and $\Delta_\mu^{(1,2)}$ are the loop corrections to the Higgs potential~\cite{Barger:1993gh}. Thus mSUGRA can be defined in terms of just three parameters and a sign: $m_{3/2}$, $m_{1/2}$, and $A_0$ plus the sign of $\mu$.

Because of the restrictive nature of electroweak symmetry breaking  in mSUGRA, the EWSB constraints can be slightly relaxed by a generalized Giudice-Masiero mechanism~\cite{ikyy,gm}, which amounts to including
\beq
\Delta K = c_H H_1 H_2  + h.c. \, ,
\label{gmk}
\eeq
in the K\"ahler potential. Here, $c_H$ is a constant and the expressions for $\mu$ and $B$ are modified at the input scale $M_{in}$. With this alteration, $\mu$ and $B$ become linearly independent, as is the case in the CMSSM,
\begin{eqnarray}
 \mu &=& \mu_0 + c_H m_{3/2}\ ,
 \label{eq:mu0}
 \\
  B\mu &=&  \mu_0 (A_0 - m_{3/2}) + 2 c_H m_{3/2}^2\ .
   \label{eq:Bmu0}
\end{eqnarray}
Above, we have maintained our assumed flat K\"ahler potential with $\mu_0$ being the $\mu$-term of the superpotential.

Now, let us move on to an even simpler version of supergravity, pure gravity mediation (PGM)\,\cite{pgm,pgm2}.
In the minimal model of PGM the K\"ahler  potential is also flat. The scalars again obtain universal scalar masses from
tree-level supersymmetry breaking effects, i.e. $m_0 = m_{3/2}$. $B$ is again related to the other supersymmetry breaking mass terms, $B_0=A_0-m_0\simeq -m_0$ (as we will see shortly $A_0 \ll m_0$ in PGM). As in the case of mSUGRA, we can free up $B_0$ by combining it with the Giudice-Masiero term~\cite{gm} given in Eq. (\ref{gmk}).
The tree-level Higgs mixing masses, $\mu$ and $B\mu$ , are then the same as those for the CMSSM which are found in Eqs.\,(\ref{eq:mu0}) and (\ref{eq:Bmu0}).

The gaugino masses and $A$-terms are, on the other hand,
assumed to have no significant sources at the tree-level.
This assumption is a natural consequence in models with no singlet supersymmetry breaking field
(i.e. the Polonyi field) or models with strong moduli stabilization\cite{klor,lmo,dlmmo}.
It is important to note that this class of models are free from the cosmological moduli/Polonyi problems.

As the tree-level gaugino masses are essentially vanishing, the dominant source for gaugino masses comes from the one-loop anomaly mediated contributions\cite{anom}, which are given
\begin{eqnarray}
    M_{1} &=&
    \frac{33}{5} \frac{g_{1}^{2}}{16 \pi^{2}}
    m_{3/2}\ ,
    \label{eq:M1} \\
    M_{2} &=&
    \frac{g_{2}^{2}}{16 \pi^{2}} m_{3/2}  \ ,
        \label{eq:M2}     \\
    M_{3} &=&  -3 \frac{g_3^2}{16\pi^2} m_{3/2}\ .
    \label{eq:M3}
\end{eqnarray}
Here, the subscripts of $M_a$, $(a=1,2,3)$, correspond to the gauge groups of the Standard Model
U(1)$_Y$, SU(2) and SU(3), respectively.
As a result, in PGM, the gaugino masses are much smaller than the scalar masses and are non-universal
even in the minimal model.
Note that at the small values of $\tan \beta$ we will consider, there are potentially large one-loop
corrections to gaugino masses.  These will be discussed in section 3 below.

In summary, the minimal model of PGM is much simpler than mSUGRA and has
only three fundamental parameters, $m_0$, $\mu$ and $B\mu$. Viable models
require a Giudice-Masiero term, $c_H$, but
using the vacuum conditions of  EWSB  in Eqs\,.(\ref{eq:mu}) and (\ref{eq:muB}), these
four parameters are further reduced to only
the two free parameters, which we may take to be
\begin{eqnarray}
 m_{3/2}, \quad \tan\beta \ .
\end{eqnarray}
In the following, we will show that $m_{3/2}$ must lie in the range of  $\mathcal{O}$(100) TeV so that
the predicted chargino mass satisfies the LEP bound, the Higgs mass
agrees with the LHC measurement, and gaugino masses
are within the reach of the LHC experiments.
With only these two parameters, as we discuss in the next section,
the model leads to a very successful phenomenology;
\begin{itemize}
\item The sfermion and  gravitino have masses $\mathcal{O}$(100) TeV.
\item The higgsino and the heavier Higgs boson also have masses $\mathcal{O}$(100) TeV.
\item The gaugino masses are in the range of  hundreds to thousands of GeV.
\item The LSP is the neutral wino which is nearly degenerate with the charged wino.
\item The lightest Higgs boson mass is consistent with the observed Higgs-like boson,
i.e. $m_h \simeq 125$--$126$\,GeV.
\end{itemize}

At first glance, it may seem difficult for models with universal scalar masses to
satisfy the vacuum conditions Eqs\,.(\ref{eq:mu}) and (\ref{eq:muB})
especially with $m_0 = {\cal O}(100)$ TeV.
As we will show, however, universal scalar masses are consistent with
electroweak symmetry breaking even for these heavy scalar mass.
In addition, we also find that the PGM model with the universal scalar mass
put an upper limit on the scalar masses of around $250$\,TeV for $m_h \simeq 125$--$126$\,GeV.

\section{Results}
Here we will discuss the parameter space of the PGM model
with the universal scalar masses.  The dominant features of PGM will be determined by radiative electroweak symmetry breaking and the perturbativity of the top Yukawa coupling.

\subsection{One-Loop Masses}
In this section, we discuss the low-scale mass spectra of PGM at the one-loop level.
This analysis will give fairly accurate approximations to the mass spectra.
As we discussed in the previous section, the mass spectrum
of the PGM model is quite hierarchical and the gaugino masses are much lighter than the scalar masses.
This hierarchy in masses drastically simplifies the one-loop RG equations.
In fact, for small $\tan\beta$, as will necessarily be the case for universal boundary conditions,
the only non-trivial beta functions are
\begin{eqnarray}
\nonumber\frac{d}{d t} m_2^2&=&\frac{3y_t^2}{8\pi^2}(m_2^2+m_{\tilde Q_3}^2 +m_{\tilde t_R}^2)\ ,\\
\frac{d}{d t} m_{\tilde Q_3}^2&=&\frac{y_t^2}{8\pi^2}(m_2^2+m_{\tilde Q_3}^2 +m_{\tilde t_R}^2)\label{eq:RGETop}\ ,\\
\nonumber \frac{d }{d t}m_{\tilde t_R}^2&=&\frac{2y_t^2}{8\pi^2}(m_2^2+m_{\tilde Q_3}^2 +m_{\tilde t_R}^2) \, ,
\end{eqnarray}
where $t = \ln Q/M_{in}$ and $Q$ is the renormalization scale.
Here, we have approximated the other third generation Yukawa couplings to be zero
since $\tan\beta$ is restricted to be rather small, as will be confirmed below.
The remaining beta functions of the MSSM are all of the order of the gaugino masses and can be neglected for this approximation.
Since the gaugino masses are loop suppressed relative to the sfermion masses,
the other sfermions effectively run at two-loops.
This effective two-loop running of the sfermions induces little change in the low-scale sfermion masses and so $m_{\tilde f}^2\simeq m_{3/2}^2$ for the other sfermions.
Due to the smallness of the other third generation Yukawa couplings for small $\tan\beta$, the low-scale
down-type Higgs soft mass is also essentially unchanged from its boundary value,
 i.e. $m_1^2 \simeq m_{3/2}^2$.

The three scalar masses, $m_2^2$, $m_{\tilde Q_3}^2$, and $m_{\tilde t_R}^2$, on the other hand,
have a real one-loop running which is proportional to the top-Yukawa coupling.
This running gives a significant deflection from their high-scale values.
The solutions to the one-loop RGE's in Eq.\,(\ref{eq:RGETop}) have a simple solution,
which is accurate to approximately
$5\%$ at low $\tan \beta$,
\begin{eqnarray}
m_2^2\simeq\frac{1}{2}m_{3/2}^2(
3I(t)-1)\ ,  \quad
m_{\tilde Q_3}^2\simeq\frac{1}{2}m_{3/2}^2
(1+I(t))\ , \quad
 m_{\tilde t_R}^2\simeq m_{3/2}^2I(t)\ ,
 \label{eq:app}
\end{eqnarray}
where
\begin{eqnarray}
I(t)=e^{\int_0^t\frac{3}{4\pi^2}y_t(t')^2 dt'} \approx \left(\frac{Q}{M_{GUT}}\right)^\frac{3 y_t^2}{4 \pi}\ ,
\end{eqnarray}
where the last approximation holds under the assumption that top quark Yukawa, $y_t$, is constant.
Interestingly, this solution obeys the sum rule, $m_2^2+m_{\tilde Q_3}^2+m_{\tilde t_R}^2=0$, in the limit
$Q\to 0$.\footnote{This sum rule is perturbed by the gluino mass, and so becomes $m_2^2+m_{\tilde Q_3}^2+m_{\tilde t_R}^2\approx m_{\tilde g}\ll m_{3/2}$.}
This is a generic feature of the RG running of PGM which persists even in non-universal cases.
In our setup, however, $y_t$ is small
and the theory never reaches its asymptotic values before
the running is stopped at $Q\simeq m_{3/2}$.

By using the above approximate solutions, we immediately find that the universal scalar mass leads
to a stringent upper bound on $\tan\beta$ for successful EWSB.
To see this, we note that when $\tan \beta$ is near its upper limit, $I(t)$ remains greater than\footnote{This is only true for larger values of $\tan\beta$. Because $I(t_{SUSY})\sim 1/3$, small changes have a drastic effect on the sign of $m_2^2$. In fact, it is regions where $I(t_{SUSY})>1/3$ that lead to the breakdown of EWSB.} $\frac{1}{3}$ during the RG evolution down to $Q\simeq m_{3/2}$, and hence, the low-scale values of $m_{1,2}^2$ are
both positive and much larger than the TeV scale.
Therefore, for successful EWSB  in Eq.\,(\ref{eq:mu}), i.e. $\mu^2 >0$,
it is possible to obtain a qualitative upper limit on $\tan\beta$ (ie., when weak scale and loop
corrections are neglected),
\begin{eqnarray}
\tan\beta \lesssim\left(\frac{m_1^2}{m_2^2}\right)^{1/2}
\simeq
\left(\frac{2}{3I(t_{\rm SUSY})-1}\right)^{1/2}\ , \quad (t_{\rm SUSY} \simeq \log m_{3/2}/M_{in}) \ .
\end{eqnarray}
For $m_{3/2}\simeq 100$\,TeV, for example, we find $I(t_{\rm SUSY}) \simeq 0.28$--$0.35$, with EWSB failing for the larger values. This then leads to an upper limit of $\tan\beta\lesssim 2$--$3$.%
\footnote{
As is evident from the above derivation, the upper limit can be relaxed
by non-universalities of the scalar masses which are
induced by a non-universal K\"ahler potential
or some running effects above the GUT scale \cite{dlmmo}.
}
Since $I(t)$ increases for larger $m_{3/2}$, we also find the upper limit on $\tan\beta$
becomes more stringent for larger $m_{3/2}$.
In the next subsection, we will confirm that $\tan\beta$ is bounded from above
in a detailed numerical calculation.

Lastly, we discuss the gaugino masses.
Gaugino masses, generated by anomaly mediation,
automatically follow the RG running, and
their masses are given by Eqs.\,(\ref{eq:M1})-(\ref{eq:M3}).
We do however include numerous one-loop corrections to these masses\,\cite{piercepapa}.
While many of these are negligibly small,
because  $\mu$ and $B \mu$ are so large,
integrating out the Higgs superfields generates threshold corrections to the wino and bino masses which are quite significant.
For example, the correction to the wino mass in this limit is approximately given by\footnote{This expression is sensitive to the sign conventions of $\mu$.}\,\cite{ggw},
\begin{eqnarray}
\Delta M_2 \simeq -\mu\frac{g^2}{16\pi^2}\sin2\beta\frac{m_A^2}{m_A^2-\mu^2}\ln\frac{m_A^2}{\mu^2}\ ,
\end{eqnarray}
where $m_A^2$ denotes the heavy Higgs pseudoscalar mass,
\begin{eqnarray}
m_A^2 = (m_1^1 + m_2^2 + 2|\mu|^2) + \Delta_A^2\ ,
\label{eq:mA}
\end{eqnarray}
with $\Delta_A^2$ being the loop corrections~\cite{Barger:1993gh}.

There are two important things to note about this expression. First is that its sign is opposite that of the anomaly mediated contribution for $\mu>0$. Secondly, it is quite similar in size to the anomaly mediated contribution.  For positive $\mu$, this contribution drastically suppresses the wino masses and for large regions of parameters space the wino masses is too small. For $\mu<0$, this contribution is positive and wino mass tends to be much heavier then that expected in anomaly mediation
as we will see in the following section. The bino mass is also corrected. However, its mass correction is much less significant due to a small numerical factor and so we do not discuss its details here.

\subsection{Parameter Scan of Pure Gravity Mediation}
We are now able to present the results of our parameter scan for PGM.
For numerical computations we employed the program {\tt SSARD}~\cite{ssard}, which uses
two-loop RGE equations for the MSSM to compute the sparticle spectrum.
We focus on the particles with masses that could possibly be seen at the LHC.
We will also examine the the relic abundance constraints on the wino LSP.

\begin{figure}[t]
\begin{minipage}{8in}
\epsfig{file=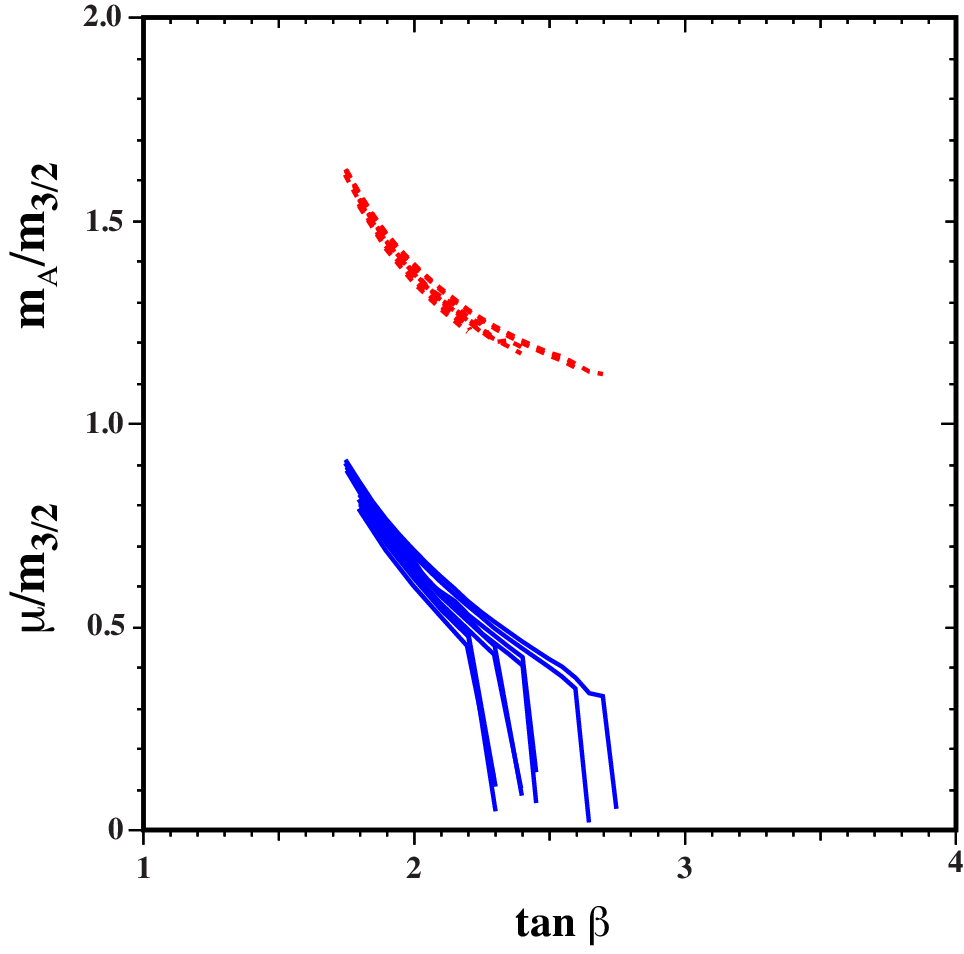,height=3.1in}
\epsfig{file=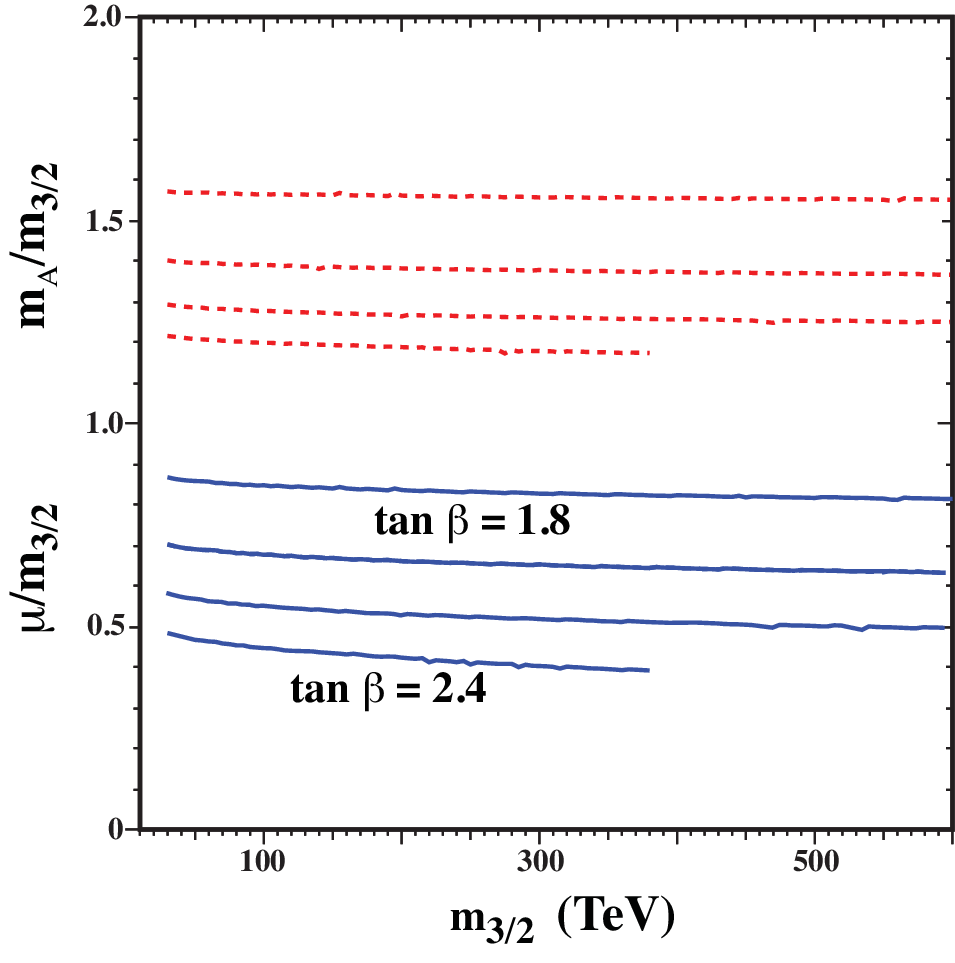,height=3.1in}
\hfill
\end{minipage}
\caption{
{\it
The Higgs mixing mass, $\mu$ (solid blue) and the Higgs pseudoscalar mass, $m_A$ (dashed red)
relative to the gravitino mass, $m_{3/2}$ as  a function for $\tan \beta$ (left) and $m_{3/2}$ (right).
In the left panel different contours correspond to different choices of $m_{3/2} = 30, 60,
100, 150, 200$, and $300$\,TeV.  The smaller values of $m_{3/2}$ extend to larger $\tan \beta$.
Notice that for each $m_{3/2}$ there is a cutoff in $\tan \beta$ where solutions
to the EWSB conditions are no longer possible. This is seen by the rapid drop in $\mu$.
In the right panel, we show four values of $\tan \beta = 1.8, 2.0, 2.2$, and $2.4$.
}}
\label{fig:muma}
\end{figure}

As was discussed previously, there are two free parameters in PGM.
We have chosen these parameters to be
\begin{eqnarray}
m_{3/2}\ , \quad \quad \tan\beta\ .
\end{eqnarray}
A priori, there is no bound on $m_{3/2}$, but for a given value of $m_{3/2}$,
the value of $\tan \beta$ is restricted to a small range, $1.7 \lesssim \tan \beta \lesssim 3$,
as shown in Fig.\,\ref{fig:muma}.
As can be seen in this figure, as $\tan\beta$ increases $\mu$ decreases.
Plotted are the values of $\mu$ and the Higgs pseudoscalar mass, $m_A$
relative to the gravitino mass.
In the left panel, we show these masses for six choices of $m_{3/2} = 30, 60, 100, 150, 200, 300$ TeV. Smaller values of $m_{3/2}$ extend to larger $\tan \beta$.
As $\tan \beta$ is increased, eventually, $|\mu|^2$ becomes negative indicating the absence
of solutions to the Higgs minimization equations and as a result, EWSB does not occur.
This is exhibited by the precipitous change in $\mu$ for larger values of $\tan \beta$.
Because of the severe fine tuning in this region, the solutions do not converge numerically and
$\mu$ is cut off prior to it going to zero.
When regular solutions are found, $\mu/m_{3/2}$ varies between about $0.3$ and $0.9$.
In the right panel, the same masses are shown as a function of the gravitino mass for four choices of $\tan \beta$ regularly spaced from $1.8$ to $2.4$.
The behavior of the pseudo-scalar mass can also be seen in Fig.\,\ref{fig:muma}. It is also larger for smaller values of $\tan\beta$ and varies between $1.2$ and $1.7\,m_{3/2}$.
Also in Fig.\,\ref{fig:muma}, we see that $\mu$ and $m_A$ roughly scale as $m_{3/2}$.
We note at this point that in all of the models considered, $c_H$ varies in the range
between -0.02-- -0.12 for $\mu > 0$ and between -0.25 -- -0.97 for $\mu < 0$.

The behaviors of $|\mu|/m_{3/2}$ and $m_A/m_{3/2}$ can be roughly understood
by using an approximate solution of Eq.\,(\ref{eq:mu})
assuming that the right hand side is dominated by the $m_1^2$  contribution
for small $\tan\beta$, and hence, we obtain
\begin{eqnarray}
\frac{|\mu|}{m_{3/2}} \simeq \left(\frac{1}{\tan^2\beta-1}\right)^{1/2}\ .
\end{eqnarray}
By plugging this result into Eq.\,(\ref{eq:mA}), we also obtain,
 \begin{eqnarray}
\frac{m_A}{m_{3/2}} \simeq \left(1+\frac{2}{\tan^2\beta-1}\right)^{1/2}\ .
\end{eqnarray}
These approximate solutions roughly reproduce the behaviors in Fig.\,\ref{fig:muma}.

\begin{figure}[h]
\begin{minipage}{8in}
\epsfig{file=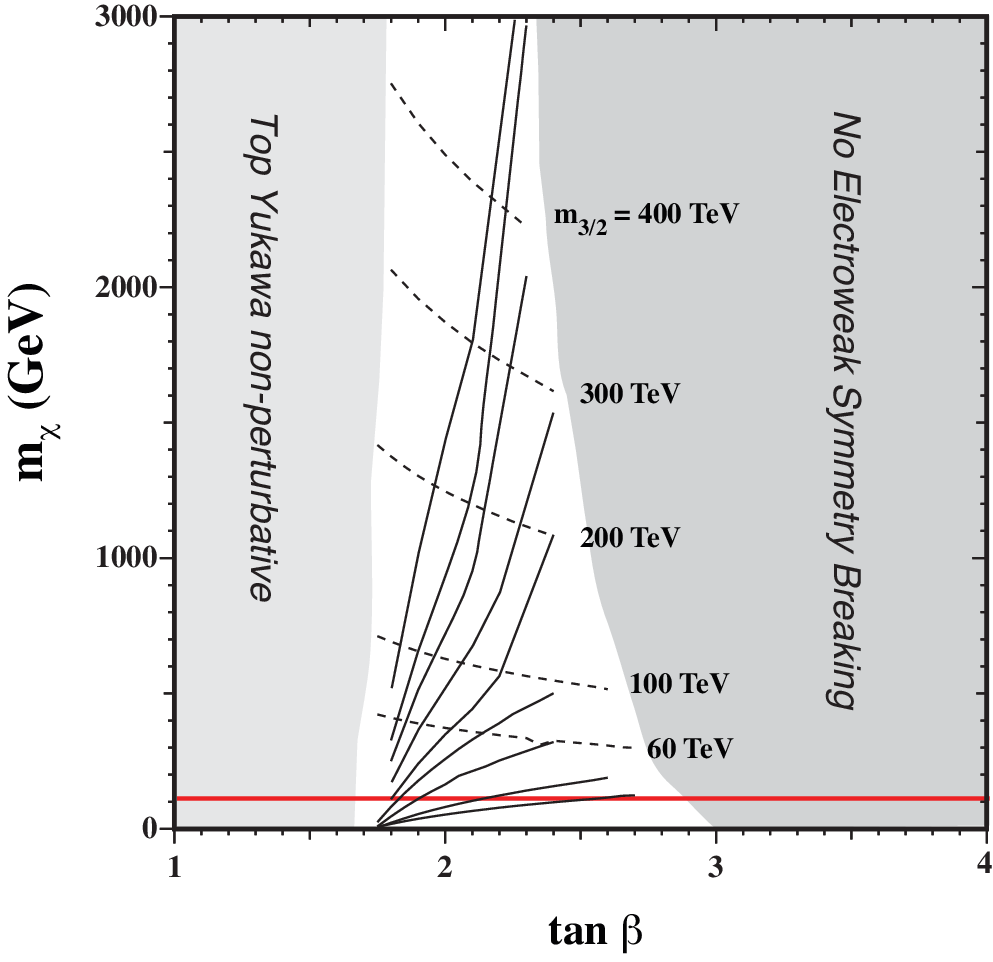,height=3.1in}
\epsfig{file=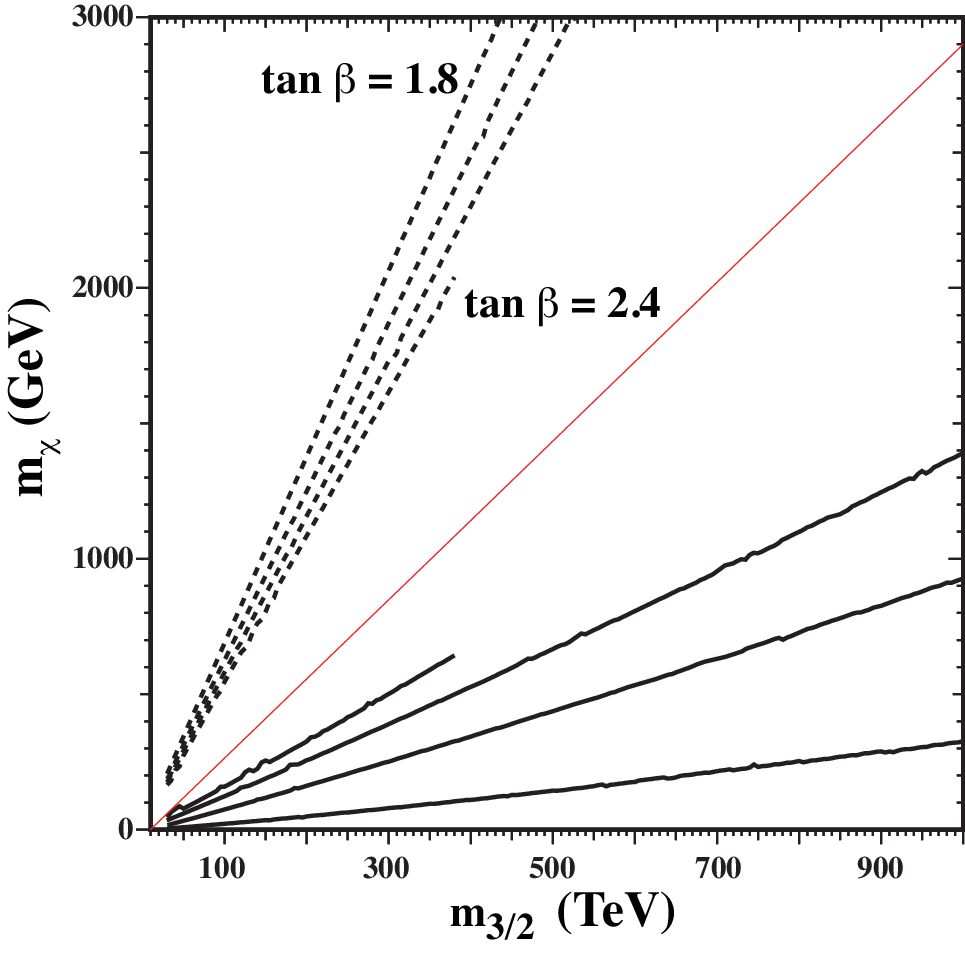,height=3.1in}
\hfill
\end{minipage}
\caption{
{\it
The wino mass as a function of $\tan \beta$ (left) and $m_{3/2}$ (right) for both
$\mu > 0$ (solid) and $\mu < 0$ (dashed).
For large values of $\tan \beta$, solutions to the EWSB conditions are not possible,
while for low values of $\tan \beta$, the top quark Yukawa diverges during the running of the RGE's.
The LEP bound of $104$\,GeV on the chargino mass is shown as the horizontal red line.
The different curves correspond to different values of $m_{3/2}$. Only the curves with $\mu < 0$
are labelled for clarity.
 In the right panel, the diagonal red line shows the wino mass when one-loop radiative corrections
 are ignored. Here the curves correspond to four values of $\tan \beta = 1.8, 2.0, 2.2$, and $2.4$.
 For $\mu > 0$, the $\tan \beta = 2.4$ curve ends when EWSB solutions are no longer possible.
}}
\label{fig:wino}
\end{figure}

\begin{figure}[h]
\begin{center}
\epsfig{file=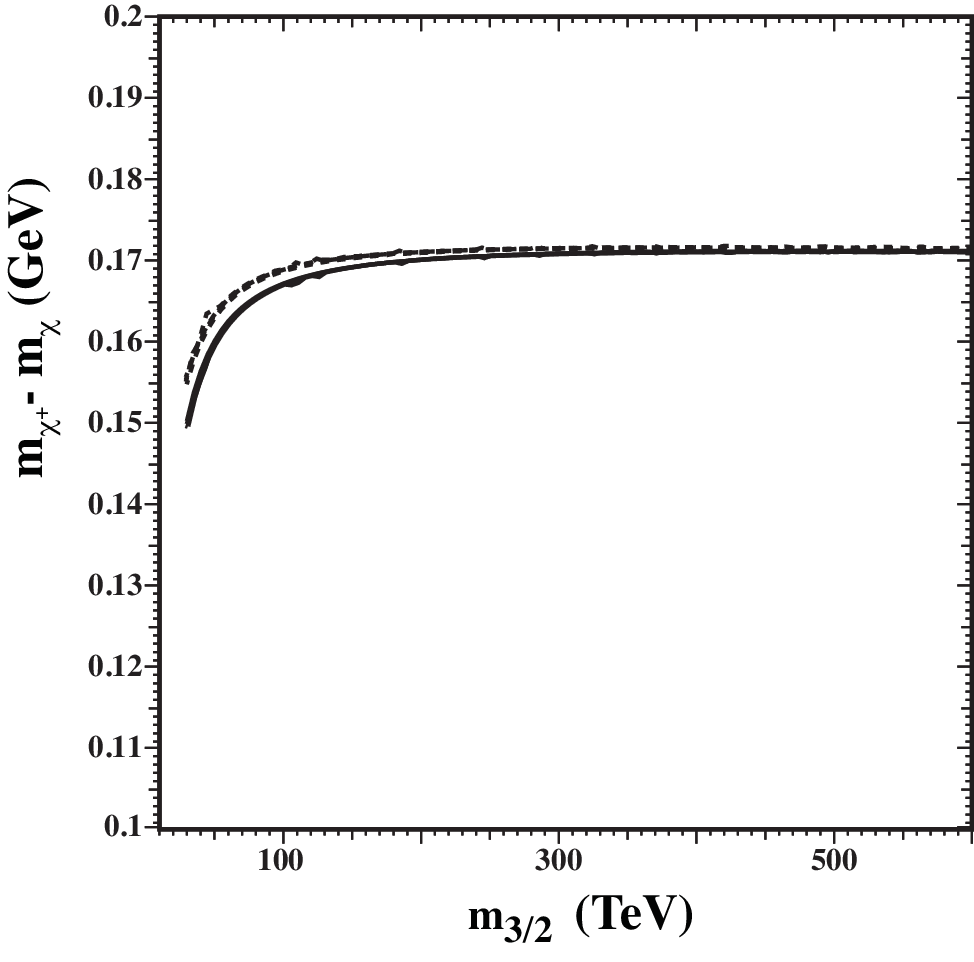,height=3.1in}
\end{center}
\caption{\it The mass different of the neutral and charged wino as a function of $m_{3/2}$.  The
solid (dashed) curve corresponds to positive (negative) $\mu$.  }\label{fig:delwino}
\end{figure}

Next, we examine the mass of the LSP.  In PGM models, the wino is the LSP.  Because of the large threshold corrections discussed above, the value of the wino mass is strongly dependent on the sign of $\mu$.  For positive values of $\mu$ and very small values of $\tan\beta$, the threshold corrections nearly cancel the anomaly mediated contribution. As $\tan\beta$ increases, the threshold corrections shrink and the wino mass approaches the anomaly mediated value. The wino mass is plotted in Fig.\,\ref{fig:wino}
as a function of $\tan \beta$  for fixed values of $m_{3/2}$ as labeled (left)
and as a function of  $m_{3/2}$ for fixed values of $\tan \beta$ (right).
Solid curves correspond to $\mu > 0$ and dashed curves to $\mu < 0$.
For clarity, only curves for $\mu < 0$ have been labeled.
The cancellation due to one-loop corrections can be seen in both panels.
For positive $\mu$, the smallest values of $\tan\beta$ are excluded by
the LEP constraint\,\cite{LEPsusy} on the charged wino mass,
where the neutral and the charged wino masses are nearly degenerate as seen in
Fig.\,\ref{fig:delwino}.%
\footnote{
See also Ref.\,\cite{ATLAS:2012jp} for
the constraints on the direct chargino production using
a disappearing-track signature at the LHC.
}
The red diagonal line in the right panel represents the anomaly mediated contribution, shown for reference. This further highlights the effect of the sign of $\mu$ on the wino mass.
As discussed previously, large values of $\tan \beta$ are not allowed because we require radiative EWSB
while perturbativity of the top quark Yukawa coupling excludes the smallest values of $\tan \beta$.
These regions are qualitatively shaded in the left panel of Fig.\,\ref{fig:wino}.

Note that so far we have assumed the absence of any new sources of CP violation.
In principle, however, we can define two new phases in universal models \cite{phase};
one associated with $A_0$ and one with $\mu$.
In the present context, the phase of $A_0$ is irrelevant.
The phase of $\mu$, on the other hand, could in fact allow us to interpolate between the
solid and dashed curves in Fig.\,\ref{fig:wino}.
It should be noted that in the above analyses, masses other than the wino mass are not altered
very much even in the presence of an arbitrary phase for $\mu$.
Furthermore, the CP-violating effects due to the phase of $\mu$
are suppressed by the heavy sfermion mass scale, and hence,
the electron electric dipole moment, for example, is predicted to lie far below the current limit
even for the CP-phases of ${\cal O}(1)$.

In Fig.\,\ref{fig:delwino}, we show the mass difference between the charged and neutral wino.
As can be seen in this figure, the charged wino is indeed heavier.%
\footnote{
For two-loop analysis on the wino mass splitting, see Ref.\cite{Ibe:2012sx}.
}
However, the mass difference is just barely larger than the pion mass.
Thus, the decay products of the charged wino will be a very soft pion and the wino LSP.
Because of this very small mass difference, the decay width will have a suppression factor of order $m_{\pi}/m_{\chi}$ and so the chargino will be quite long lived.
The experimental signatures of this decay have been discussed in PGM\,\cite{pgm,pgm2} and related scenarios \cite{ggw,Feng:1999fu,ctchar,dlmmo}.

\begin{figure}[h]
\vskip 0.5in
\vspace*{-0.45in}
\begin{minipage}{8in}
\epsfig{file=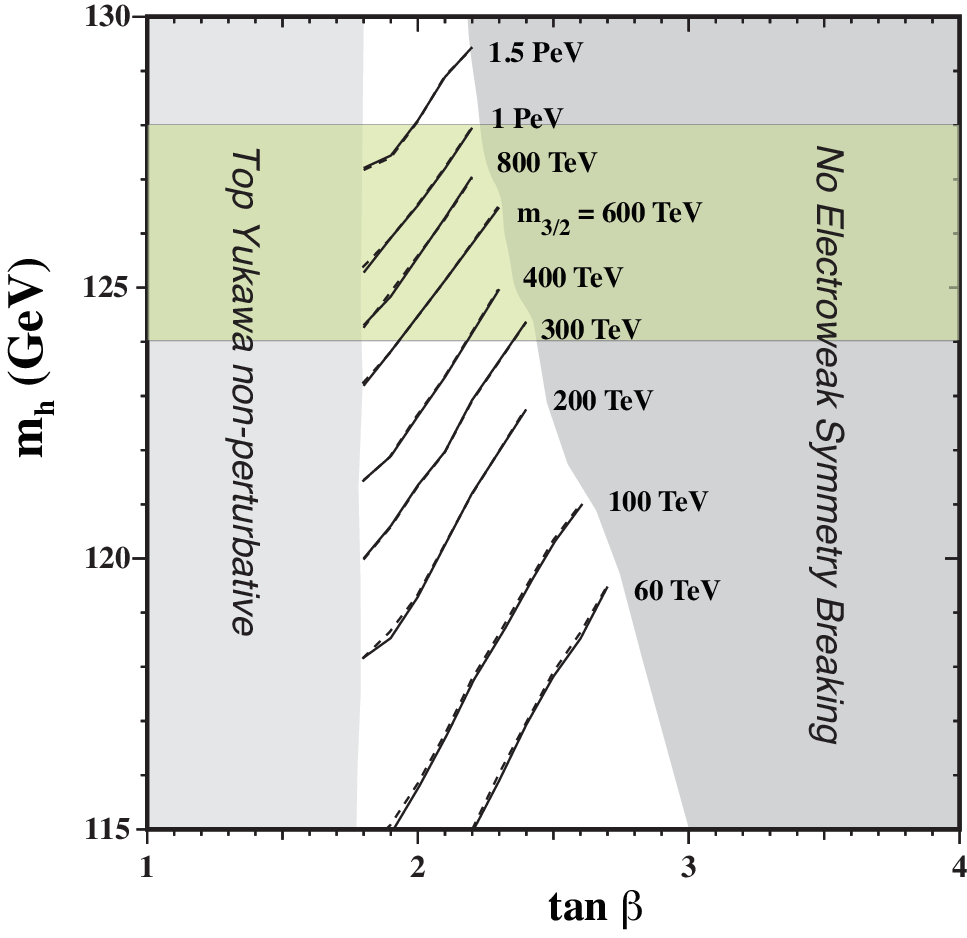,height=3.1in}
\epsfig{file=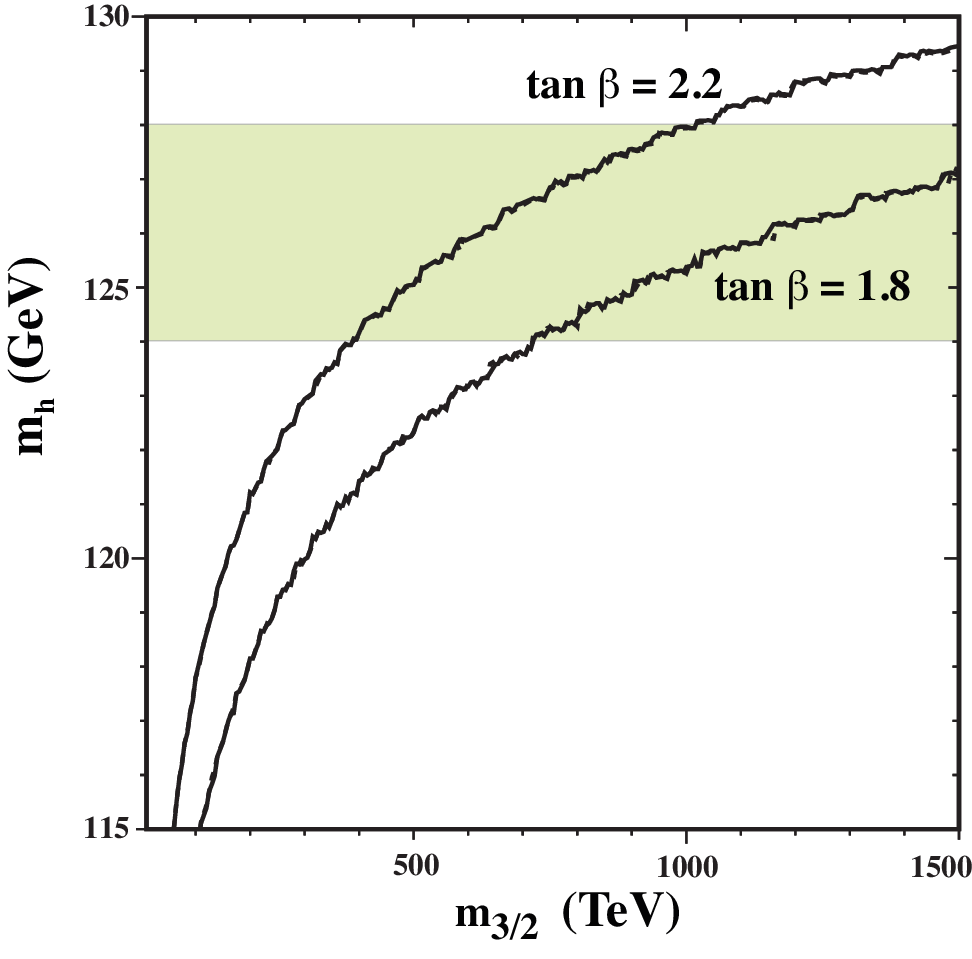,height=3.1in}
\hfill
\end{minipage}
\caption{
{\it
The light Higgs mass as a function of $\tan \beta$ (left) and $m_{3/2}$ (right) for both
$\mu > 0$ (solid) and $\mu < 0$ (dashed).  For large values of $\tan \beta$,
solutions to the EWSB conditions are not possible, while for low values of $\tan \beta$, the top
quark Yukawa diverges during the running of the RGE's.
The LHC range (including an estimate of uncertainties) of $m_h = 126 \pm 2$ GeV
is shown as the pale green horizontal band.
The different curves correspond to different values of $m_{3/2}$ as marked.
In the right panel, the curves correspond to two values of $\tan \beta = 1.8 $and $2.2$.
}}
\label{fig:mh}
\end{figure}

Next, we examine the predictions for the Higgs mass.
The calculation here follows \cite{dlmmo} and is based on derivations found in \cite{mhsplit}.
In Fig.\,\ref{fig:mh}, we show the Higgs mass as a function of $\tan \beta$
for fixed values of $m_{3/2}$ (left)  and as a function of $m_{3/2}$ for fixed values of $\tan \beta$ (right). There is little dependence on the sign of $\mu$ as seen by the similarity of the solid ($\mu > 0$) and dashed ($\mu < 0$) curves.
The horizontal band spans $m_h = 126 \pm 2$ GeV shown to guide the eye.
Here, and in all of the numerical work presented, we have assumed a top quark mass of
$m_t = 173.2$ GeV.
The experimental uncertainty of $\pm 0.9$ GeV\,\cite{mt} would translate into a shift in
$m_h$ of  roughly 1\,GeV.
In the left panel, we again show qualitatively the regions at low and high $\tan \beta$
which are excluded due to  top quark Yukawa perturbativity and successful EWSB, respectively.

Because $A$-terms are so small, the Higgs mass is predominantly enhanced by stop corrections
to the quartic coupling in most of the parameter space.
Since these corrections grow logarithmically with the stop masses,
the Higgs boson mass also increases with $m_{3/2}$ as can be see in Fig.\,\ref{fig:mh}.
It should also be noted that the stop mixing parameter
$X_t= A_t+\mu\cot\beta$ becomes large at the low end of $\tan\beta$
where $\mu$ is rather large as discussed above.
Thus, in that region, the stop mixing effect to the one-loop Higgs
potential slightly enhances the Higgs boson mass,
which can be seen in the left panel of Fig.\,\ref{fig:mh} as a slight increase of $m_h$.
As $\tan\beta$ is increased, $\mu$ decreases, and hence, the significance of the  $X_t$ contribution
to the Higgs mass disappears and the Higgs mass decreases slightly.
Increasing $\tan\beta$ further, increases the Higgs boson mass.
This is because the tree-level contribution to the Higgs mass is proportional to $\cos^22\beta$.
Despite these rather involved dependences on $\tan\beta$,
the Higgs mass remains fairly constant for a given value of $m_{3/2}$ due to the restrictions in the allowed
range of $\tan\beta$.
Examining these plots, we see that experimental bounds on the Higgs mass constrain the gravitino mass to be in the range $m_{3/2}=300$--$1500$ TeV.

\begin{figure}[h]
\vskip 0.5in
\vspace*{-0.45in}
\begin{minipage}{8in}
\epsfig{file=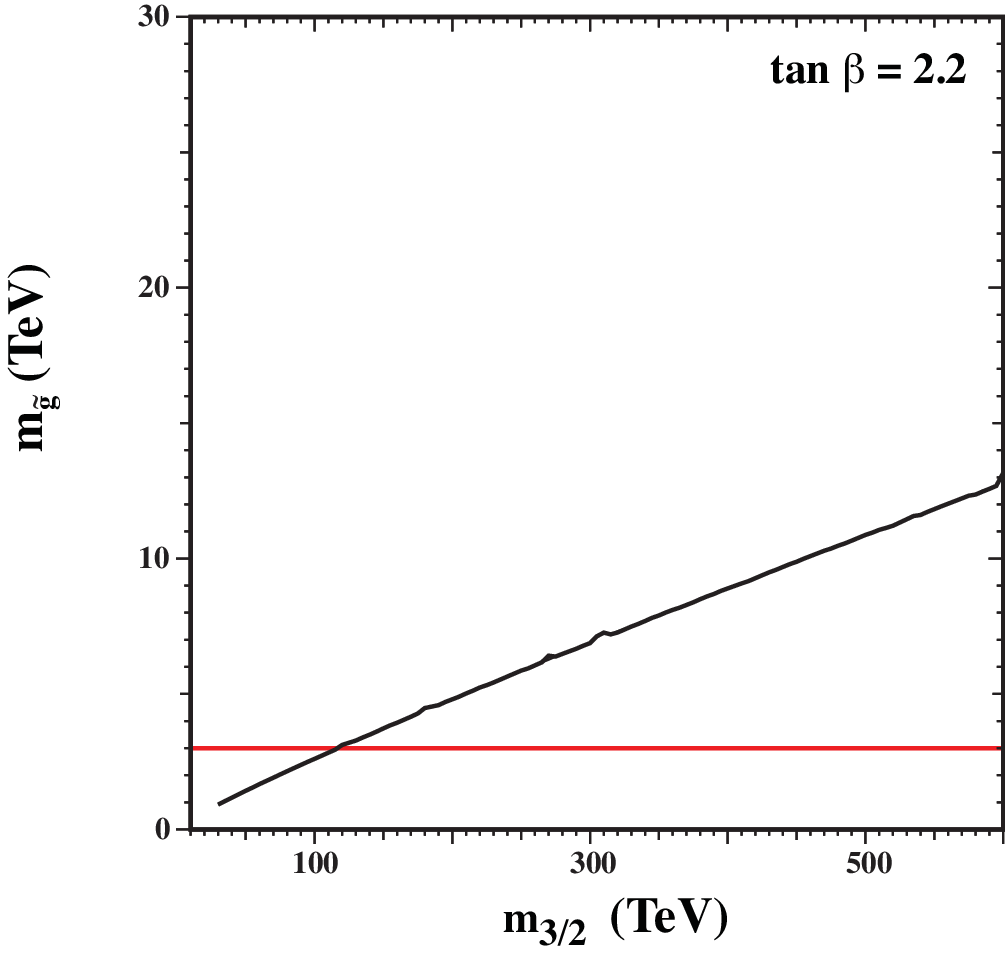,height=3.1in}
\epsfig{file=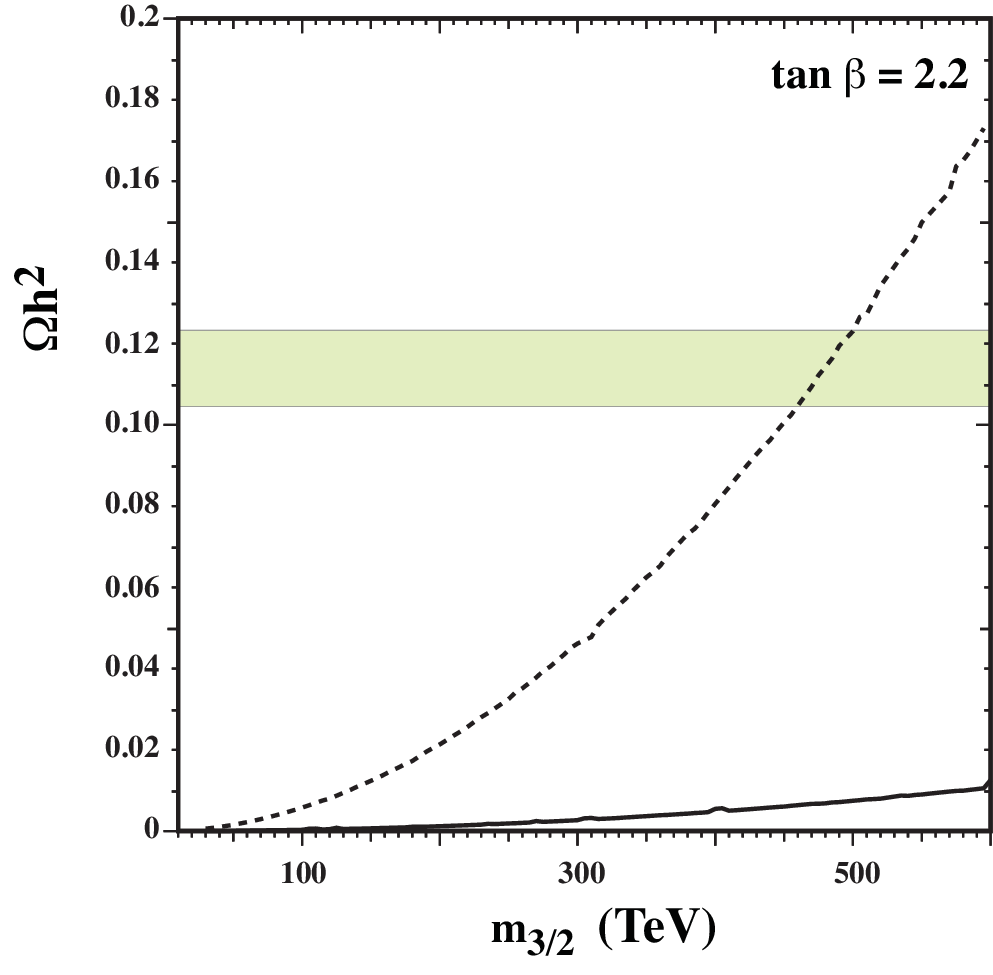,height=3.1in}
\hfill
\end{minipage}
\caption{
{\it
The gluino mass (left) and relic density, $\Omega_\chi h^2$ (right),
as a function of $m_{3/2}$ (right) for both
$\mu > 0$ (solid) and $\mu < 0$ (negative). The gluino mass is not sensitive to the sign of $\mu$ and the two curves lie on top of each other.
 In the right panel, the WMAP range for the relic density is shown by the pale green horizontal band.
 The relic density crosses this band for $\mu <0$ when $m_{3/2} \simeq 450$--$500$\,TeV, however,
 at these gravitino masses the Higgs mass is in excess of $130$\,GeV as seen in Fig.~\protect{\ref{fig:mh}}.
}}
\label{fig:mgohsq}
\end{figure}

In Fig.\,\ref{fig:mgohsq}, we show the gluino mass, $m_{\tilde g}$ as a function of $m_{3/2}$.
The gluino mass is essentially independent of $\tan \beta$, and we fixed $\tan \beta = 2.2$.
As one might expect, the gluino mass is effectively given by Eq.\,(\ref{eq:M3})
(any deviation is due to two-loop effects which are included in the running).
As summarized in Ref.\,\cite{pgm2}, the most severe limit on the gluino mass
is obtained from the search for multi-jets plus missing transverse energy events
at the LHC, which leads to $m_{\tilde g} > 1.2\,(1.0)$\,TeV
for a wino mass of $100\,(500)\,$GeV \cite{atlas}.%
\footnote{
As we have discussed in the previous section, the universal boundary condition
leads to the lighter third generation squarks than the other squarks.
In this case, the gluino tends to decay into a wino with a pair of top quarks and bottom quarks.
Thus, the final state of the gluino production events include many $b$-quarks which
slightly change the sensitivities to multi-jets plus missing transverse energy events.
(See also Ref.\,\cite{pgm2}.)
}
The red line represents the approximate  reach of the LHC.
A gluino mass of roughly $3$\,TeV corresponds to a gravitino mass of $m_{3/2}=110$\,TeV.
A gravitino mass this small is excluded if the Higgs mass is required to be larger than $124$ GeV.
This means the only hope for probing universal PGM models at the LHC is through the lightest chargino.

Finally, we show the dependence of the wino relic density, $\Omega_\chi h^2$ in the right panel of
Fig.\,\ref{fig:mgohsq} as a function of $m_{3/2}$.
The relic density is also essentially independent of $\tan \beta$, and we fixed $\tan \beta = 2.2$.
As one can see, the relic density is always extremely small for $\mu > 0$
and therefore dark matter must either come from some other sector of the theory (e.g. axions)
or there must be a non-thermal source for wino production after freeze-out\footnote{For a wino mass less than $500$ GeV, the wino constituent of dark matter is restricted to be a fraction of the total dark matter\cite{Hall:2012zp}.}.
For $\mu < 0$, it is possible to get an acceptable relic density when $m_{3/2}\simeq 460 - 500$\,TeV.
Amazingly, this is consistent with the Higgs mass measurements if the ratio of the Higgs vevs is in the rather small range $\tan\beta =2.1-2.3$.
Thus, while an alternate source for dark matter is needed for $\mu > 0$, thermal dark matter is possible for  $\mu < 0$.

\section{Conclusion}
In the pre-LHC era, the minimal models of supergravity, mSUGRA, having three parameters
and the closely related CMSSM having four could satisfy all experimental constraints.
Indeed, they offered a significantly better fit to low energy precision data than the
Standard Model\,\cite{mc}.
However, parameters such as $m_{1/2}$ and $A_0$ were included in order to preserve a notion of  naturalness with $m_{0}\sim m_{1/2}$, and supersymmetry needed to be broken by a singlet such
as the Polonyi field.
If supersymmetry is broken by a singlet, gauginos can be coupled to supersymmetry breaking
at the tree-level through the gauge kinetic function with some arbitrary coefficient.
This arbitrary coefficient added freedom making the gaugino masses universal but independent of the scalar masses.
With a singlet breaking SUSY, $A$-terms would also be generated at tree-level.

After the initial run of the LHC, the most stringent definitions of naturalness seem to be incorrect.
For generic models, the superpartners are required to be larger than about $1$\,TeV.
 Furthermore, Atlas and CMS have seen a resonance at around $125$\,GeV,
 which is more than likely the Higgs boson.
 If this is indeed the case, most supersymmetric models need large stop masses to explain
 this Higgs boson mass.
 However, in mSUGRA and the CMSSM, one can not take arbitrarily large values
 of the mass parameters and still satisfy EWSB and the upper limit on the relic density
simultaneously except in some extremely narrow regions of the parameter space.

One possibility for realizing a heavy Higgs boson is, in fact,  to take a more minimal model of supergravity, pure gravity mediation (PGM). In this model, supersymmetry is not broken by a singlet.
With no singlet to couple to the gauginos, their mass is generated at one-loop from anomalies and will be much less than the scalar masses, $m_{1/2}\ll m_{0}$.
To meet experimental constraints on the gauginos masses, their masses must be somewhat larger
than the weak scale. This means that the scalar masses will be a loop factor larger than the gaugino masses.
Seeming to be much more than a coincidence, this coincides with the mass scale needed to explain
a $125$ GeV Higgs boson mass\,\cite{oyy}.
However, just as EWSB is highly restrictive in mSUGRA, so it is in the minimal model of PGM
and there is no acceptable EWSB.
These restrictions on EWSB can be loosened in an identical manner as the restriction in mSUGRA
can be loosened to give the CMSSM.
With the addition of a Giudice-Masiero term to the K\"ahler potential, a viable model with two parameters, $m_0=m_{3/2}$ and $\tan\beta$, and acceptable EWSB can be achieved.

In this two parameter model, successful EWSB and perturbative Yukawa couplings all the way up to the GUT scale, restricts $\tan\beta$ to be in a fairly narrow range $\tan\beta =1.7$--$3$.
The gravitino mass, on the other hand, is not restricted by either of these constraints.
Its sole constraint comes from requiring the Higgs mass to be around $125$\,GeV.
To get a Higgs mass in the LHC range, the gravitino mass must fall between
$m_{3/2}=300$--$1500$\,GeV.
For this mass range electroweak gauginos may still be accessible at the LHC.
Because of large threshold corrections to the wino masses from the Higgsinos,
the wino masses are quite sensitive to the sign (or the phase) of $\mu$.
For positive $\mu$, this correction is negative and the wino mass is strongly suppressed.
In which case, the wino would have a very good chance of being seen at the LHC.
For negative $\mu$, the wino mass is much heavier and could still be seen at the LHC,
but with much more difficulty. As in most models of anomaly generated gaugino masses,
the chargino and neutral wino are nearly degenerate in mass and this too offers
a potential experimental signature.

Our final remark on PGM models with the universal scalar masses
is about the possibility of thermal wino
dark matter.
The range for the gravitino mass is limited by the necessity of obtaining a Higgs boson mass
in the LHC range, which leads to a wino LSP which is too light to yield a thermal relic density large
enough to be the dominate source of dark matter when $\mu > 0$.
In this case, wino dark matter must arise non-thermally after freeze-out
(e.g. by the decay of the gravitino or a moduli), or an additional dark matter candidate is needed.
However, for $\mu < 0$,
one finds large postive corrections to the wino mass such that for $m_{3/2}$ in
the range of 460 -- 500 GeV, one obtains the correct thermal abundance of dark matter.

\section*{Acknowledgments}
We would like to thank A. Mustafayev for helpful conversations. We would also like to thank T. Moroi and M. Nagai for helping us correct our Higgs mass calculation. 
The work of J.E. and K.A.O. was supported in part
by DOE grant DE--FG02--94ER--40823 at the University of Minnesota.
This work is also supported by Grant-in-Aid for Scientific research from the
Ministry of Education, Science, Sports, and Culture (MEXT), Japan, No.\ 22244021 (T.T.Y.),
No.\ 24740151 (M.I.), and also by the World Premier International Research Center Initiative (WPI Initiative), MEXT, Japan.

\end{document}